\documentstyle[12pt,epsfig]{article}

\def\bib{\bibitem}
\def\be{\begin{equation}}
\def\ee{\end{equation}}
\def\barr{\begin{array}}
\def\earr{\end{array}}

\def\beq{\begin{eqnarray}}
\def\eeq{\end{eqnarray}}
\def\ra{\rightarrow}

\def\etal{ {\em et al.}}

\def\lsim{\:\raisebox{-0.5ex}{$\stackrel{\textstyle<}{\sim}$}\:}
\def\gsim{\:\raisebox{-0.5ex}{$\stackrel{\textstyle>}{\sim}$}\:}

\def\ra{\rightarrow}

\def\e+e-{$e^+e^-$}
\def\snu{\tilde{\nu}}

\def\ib#1,#2,#3{       {\it ibid.\/ }{\bf #1} (19#2) #3}
\def\ap#1,#2,#3{       {\it Ann.~Phys.~(NY)\/ }{\bf #1} (19#2) #3}
\def\ijmp#1,#2,#3{     {\it Int.~J.~Mod.~Phys.\/ } {\bf A#1} (19#2) #3}
\def\mpl#1,#2,#3 {     {\it Mod.~Phys.~Lett.\/ } {\bf A#1} (19#2) #3}
\def\np#1,#2,#3{       {\it Nucl.~Phys.\/ }{\bf B#1} (19#2) #3}
\def\npps#1,#2,#3{     {\it Nucl.~Phys.~B (Proc.~Suppl.)\/ }{\bf B#1}
                             (19#2) #3}
\def\plb#1,#2,#3{      {\it Phys.~Lett.\/ }{\bf B#1} (19#2) #3}
\def\pr#1,#2,#3{       {\it Phys.~Rev.\/ }{\bf #1} (19#2) #3}
\def\prd#1,#2,#3{      {\it Phys.~Rev.\/ }{\bf D#1} (19#2) #3}
\def\prep#1,#2,#3{     {\it Phys.~Rep.\/ }{\bf #1} (19#2) #3}
\def\prl#1,#2,#3{      {\it Phys.~Rev.~Lett.\/ }{\bf #1} (19#2) #3}
\def\pro#1,#2,#3{      {\it Prog.~Theor.~Phys.\/ }{\bf #1} (19#2) #3}
\def\rmp#1,#2,#3{      {\it Rev.~Mod.~Phys.\/ }{\bf #1} (19#2) #3}
\def\sp#1,#2,#3{       {\it Sov.~Phys.-Usp.\/ }{\bf #1} (19#2) #3}
\def\zpc#1,#2,#3{      {\it Zeit.~f\"ur Physik\/ }{\bf C#1} (19#2) #3}
\def\appb#1,#2,#3{     {\it Acta Phys.\ Polon.\/ }{\bf B#1} (19#2) #3}

\begin{document}
\thispagestyle{empty}
\setcounter{page}{0}
\renewcommand{\thefootnote}{\fnsymbol{footnote}}

\begin{flushright}
DESY 96-255\\[1.7ex]
{\large \tt hep-ph/yymmddd} \\
\end{flushright}

\vskip 45pt
\begin{center}
{\Large \bf Decays of $W$ Bosons to Charginos and Neutralinos}

\vspace{11mm}
{\large J. Kalinowski}$^{1,2}$, 
{\large  and P.M. Zerwas}$^1$\\[1.1ex]
{\em $^1$ Deutches Elektronen Sychrotron DESY, D-22607 Hamburg,  
Germany.}\\[1.1ex]
{\em $^2$ Institute of Theoretical Physics, Warsaw University, 
PL-00681 Warsaw, Poland}
\\[2ex]


\vspace{50pt}
{\bf ABSTRACT}
\end{center}
\begin{quotation}
The region of the supersymmetry parameter space, in which charginos 
decay predominantly into sneutrinos and leptons: $\chi^+_1 \ra \tilde{\nu} 
+ l^+$, is not excluded experimentally for small mass differences 
between charginos and sneutrinos. The decay sneutrinos are invisible 
in R-parity conserving theories since they are either the lightest 
supersymmetric particles or  they decay primarily into the channel 
$\tilde{\nu} \ra \nu+\chi^0_1$. If the energy of the decay leptons 
is so small that they escape detection, chargino events $e^+e^- \ra 
\chi^+_1\chi^-_1$ in $e^+e^-$ collisions remain invisible, eroding 
the excluded chargino mass range at LEP. This region of the supersymmetry 
parameter space can partly be covered by searching for single $W$ 
events in $e^+e^- \ra W^+W^-$, with one $W$ boson decaying to leptons 
or quark jets, but the second $W$ boson decaying to 
(undetected) charginos and neutralinos.
\end{quotation}

\newpage
\renewcommand{\thefootnote}{\arabic{footnote}}
1. The \e+e- collider LEP is an ideal instrument to search for 
charginos and neutralinos in supersymmetric theories. 
These particles  are mixtures of gauginos and 
higgsinos, the supersymmetric partners of gauge bosons and Higgs bosons 
\cite{R1}. 
They can be  produced pairwise in \e+e- collisions: 
\beq
e^+e^- &\ra& \chi^+_i \chi^-_j \hbox{~~~for~~~} i,j=1,2 \\
e^+e^- &\ra& \chi^0_i \chi^0_j \hbox{~~~~for~~~} i,j=1,...,4 
\eeq
If the sneutrinos $\snu_{eL}$ are heavy, charginos can be probed up to the 
kinematical limit; if they are light, the upper bound 
on the chargino mass is reduced 
dramatically for gaugino-like charginos  by the destructive 
interference between the s-channel 
$\gamma$, $Z$ and the t-channel $\snu_{eL}$ exchange 
diagrams in the production amplitude.

However, within low-energy supersymmetry there is a small spot in the
SUSY parameter space in which the experimental search technique
\cite{R2} for the lightest chargino in the production process
$e^+e^-\ra \chi^+_1\chi^-_1$ fails.  If one of the sneutrino masses is
just below the $\chi^+_1$ mass, $m(\snu_{lL})\lsim m(\chi^+_1)$, the
dominant decay mode is the 2-body decay $\chi^+_1 \ra \snu_{lL} l^+$,
since the associated left-chiral slepton $\tilde{l}_L$ is too heavy
for the other 2-body decay channel $\chi^+_1\ra \tilde{l}_L\nu_l$ to
be open.  If the mass gap is narrow, the decay lepton $l^+$ is emitted
with very low energy in the $\chi^+_1$ rest frame.  The decay
sneutrino is invisible since it is either the lightest supersymmetric
particle or it decays primarily to the associated neutrino and the
lightest neutralino $\chi^0_1$.  As a result, chargino events in
$e^+e^-$ collisions are invisible because the energy of the decay
leptons is so small that they escape detection. In such a
case, called "blind spot" in Ref.\cite{R2}, charginos can be as light
as 45 GeV, the ultimate limit obtained in the search for this particle
at LEP1 \cite{R2a}.

Several methods can be used  to eliminate this spot. Constraints 
from future high-precision $(g-2)_{\mu}$  measurements have been 
exploited  in 
Ref.\cite{R3}; this method is successful for large $\tan\beta$.

However, the LEP experiments themselves can probe this exceptional
part of the supersymmetry parameter space in several ways.  (i) If the
energy is sufficiently above the chargino mass, the boost of the decay
lepton becomes so large that the particle can be tracked in the final
stale. (ii) The region $m(\snu_{lL}) \lsim m(\chi^+_0)$ in the
parameter space can be excluded if the associated left-chiral slepton
with a mass $m^2(\tilde{l}_L)=m^2(\snu_{lL})+ \cos^2\theta_W |\cos
2\beta| M_Z^2$ is not observed. (iii) The annihilation event may be
tagged by observing single photons in $e^+e^- \ra
\gamma+\chi^+_1\chi^-_1$, with the two charginos not detected.  The
upper limit of the $\gamma$ spectrum, $E_{\gamma}\le
[s-4m^2(\chi^+_1)]/2\sqrt{s}$, is characteristic for the production of
the massive chargino pairs. The cross section for this photonic
reaction however is very small and the background due to radiative $Z$
return is large. (iv) Finally, the production of $WW$ pairs can be
used to explore this parameter region. If one of the two $W$ bosons
decays undetected into a chargino plus neutralino pair, 
\beq \nonumber e^+e^-
&\ra& W^+W^-\\ & & \hspace{2mm} \raisebox{1.05ex}{$|$}\hspace{-6pt}
\longrightarrow \chi^+\chi^0 
\eeq 
"single $W$" final states,
$e^+e^-\ra W $ + (no other visible particle), are generated by this
mode.  In essence, the SM decays of the $W$ boson in one hemisphere
are used to track down invisible decays of the second $W$ in the
opposite hemisphere.  This method is best suited for two real $W$
bosons with $m(W^+) > m(\chi^+_1) + m(\chi^0_1)$; if this condition is
not fulfilled any more, the decay $W$ boson is virtual and the rate is
reduced significantly.  We find that if such nonstandard invisible $W$
boson decays occur at the level of a few percent, they should be
detectable at LEP energies.  Thus, the non-observation of "single $W$"
events could therefore be used to close the blind spot in
supersymmetry parameter space.

2. In this letter we investigate the decay of the $W$ bosons 
to charginos and neutralinos:
\beq W^+\ra \chi^{\pm}_i\chi^0_j \mbox{~~~~~~}[i=1,2;\, j=1,...,4]
       \label{wdec}
\eeq
In practice, we can restrict ourselves to the lightest chargino 
to allow for maximum phase space, though the generalization to  
the other pairings is straightforward. 
In some areas of the parameter space the heavier neutralinos 
$\chi^0_j$ may still be light enough and  their coupling 
large enough to allow for $W$ decays into these states too. 
In the numerical analysis all kinematically possible decay modes 
to charginos and neutralinos 
will be taken into account.

The analysis is set up within the frame of the minimal low-energy
supersymmetric theory. Even though no reference is made to underlying
unified theories, we assume the  relation
$M_1=\frac{5}{3}M_2\,\tan^2\theta_W$ for gaugino mass parameters 
just for sake of simplicity; this
assumption can easily be lifted.  Within this frame, the
chargino/neutralino sector is characterized by three parameters: the
mixing angle $\beta$, the wino mass $M_2$ and the higgsino mass
parameter $\mu$.

Extending the calculations of Refs.\cite{R5} to the case of general
mixing in the chargino and neutralino sectors, the partial widths for
the decay processes (\ref{wdec}) are given by the expressions \beq
&&\Gamma (W^+\ra \chi^+_i\chi^0_j)= \frac{G_F m^3_W
  \lambda_{ij}^{1/2}} {6\sqrt{2}\pi}
   \\
  &&\times \left\{ \left[2-\kappa^2_i-\kappa^2_j
    -(\kappa^2_i-\kappa^2_j)^2\right] 
(Q^2_{Lij}+Q^2_{Rij})+
  12\kappa_i \kappa_j\, Q_{Lij}\, Q_{Rij}\right \} \nonumber
\eeq
where, with  $m_{i,j}$ being the  chargino/neutralino masses,  
$\kappa_i=m_i/m_W$ and  $\lambda_{ij}=
(1-\kappa^2_i-\kappa^2_j)^2-4\kappa^2_i \kappa^2_j$, the usual 2-body 
phase space coefficient. 
The couplings of the $W$ boson 
to charginos and neutralinos are written in the usual form as
\beq
Q_{Lij}&=&  Z_{j2} V_{i1} - \frac{1}{\sqrt{2}} Z_{j4} V_{i2}  \\
Q_{Rij}&=&  Z_{j2} U_{i1} + \frac{1}{\sqrt{2}} Z_{j3} U_{i2}
\eeq
where $U$, $V$ are the mixing matrices in the  chargino sector, and 
$Z$ in neutralino sector\cite{R5a}.

Since the $W$ bosons are generated predominantly in a state of
transverse polarization, the decay to the $\chi^+\chi^0$ pair is not
isotropic.  However, the impact of the $W$ polarization on the angular
distribution is small due to the large masses of the decay particles.
We have quantitatively checked this point by analyzing the angular
distribution of polarized $W$ decays.  The spin axis is chosen as the
reference axis for the polar angle $\theta$ of the chargino/neutralino
axis. In this frame the angular distribution for the decay
$W^{\pm}(S_z=\pm 1,0)\ra\chi^{\pm}_i\chi^0_j$ is given by \beq
\frac{\mbox{d}\Gamma^{\pm}}{\mbox{d}\cos\theta} &\sim&
[1-\kappa^2_i-\kappa^2_j-\frac{1}{2}\lambda_{ij}
\sin^2\theta]\,[Q^2_{Lij}+Q^2_{Rij}]
+4\kappa_i\kappa_jQ_{Lij}Q_{Rij}\\ 
\frac{\mbox{d}\Gamma^{0}}{\mbox{d}\cos\theta} &\sim&
[1-\kappa^2_i-\kappa^2_j-\lambda_{ij}\cos^2\theta]\,[Q^2_{Lij}+Q^2_{Rij}]
+4\kappa_i\kappa_jQ_{Lij}Q_{Rij} \eeq Since the $W\chi\chi$ coupling
is ${\cal C}$ and ${\cal P}$ symmetric, no forward--backward asymmetry
is generated.

The range of the parameters $M_2$ and $\mu$, which determine masses
and couplings of charginos and neutralinos, is restricted by the $Z$
decays at LEP1 \cite{R2a}.  The impact on the $\chi^0_1$ mass from the
Tristan analysis \cite{R7} of the selectron production for
$m(\tilde{e}) \gsim 65$ GeV is small.  From the LEP runs above the
$Z$, the parameters are restricted by the non-observation of
$\chi^0_1\chi^0_i$ signals \cite{R8}.  Note however that these bounds
have not been derived for the special condition $m(\snu_{lL})\lsim
m(\chi^+_1)$ which the present analysis is based upon.  They should
therefore be taken only as a general guideline.  The measurement of
the $W$ width and the $W$ decay branching ratios at the Tevatron
\cite{R9} provide additional restrictions on possible supersymmetric
decay modes of the $W$ boson; we demand that
$\Gamma(W\ra\chi\chi)<140$ MeV be fulfilled for the $\chi$ masses
discussed in this note. The envelope of the constraints listed above
is shown by the broken lines in Fig.1 for the representative value
$\tan\beta=1.5$. The zone excluded is the area below and in between
these lines.  For large $\tan\beta$, the $\chi^+_1$ and $\chi^0_1$
mass constraints forbid on-shell $W$ decays to charginos and
neutralinos.

The main result of this letter is summarized in Fig.1. The solid
contour lines represent $W$ decay branching ratios of 7, 5, 3 and 1\%
into chargino and neutralino final states within the area allowed by
the mass constraints for $\tan\beta=1.5$. The only channels which
enter the total width in the $W$ branching ratios, are the usual
channels of the Standard Model and the new $\chi^{\pm}_i\chi^0_j$
decay channels. The numbers quoted above for the branching ratios
correspond to partial widths of approximately 140 MeV down to 20 MeV.
An additional narrow strip is excluded for positive $\mu$ by the
boundary condition BR$(W\ra \chi\chi)<7$\%. For better illustration,
the contour lines are shown in Fig.2 only in the window $100< \mu<300$
GeV and $60<M_2<140$ GeV. The contour lines continue in narrow strips
to larger $\mu$ and $M_2$ values.  {}For negative $\mu$, $W$ decays to
charginos and neutralinos can occur in the strip $\mu<-20$ GeV
adjacent to the LEP1 limit.

The same contour lines for the $W$ branching ratios of 7, 5, 3 and 1\% 
are shown in Fig.3  as a function of $[m(\chi_1^+),m(\chi^0_1]$,
separately for positive and negative $\mu$.  Only solutions with
$m(\chi^0_2)>45$ GeV are displayed.  The regions corresponding to
higgsino-like (large $M_2$) or gaugino-like (large $|\mu|$) light
charginos and neutralinos, are marked explicitly.  For positive $\mu$
the contour lines extend up to $m(\chi_1^+)\sim 54$ GeV, for negative
$\mu$ up to $m(\chi_1^+)\sim 65$ GeV. In the case of negative $\mu$,
there are two branches of the contour lines for 1 and 3\%, in analogy
to Fig.1.  Some contour lines terminate inside the figure because
either the $m(\chi^0_2)=45$ GeV limit is reached or one of the $M_2$
or $|\mu|$ parameters is larger than 400 GeV.

3. From the contour lines given in Figs.1/3, it is manifest that
$W\ra\chi\chi$ branching ratios up to order 7\% are still in the
allowed zones of the $[m(\chi_1^+),m(\chi^0_1)]$ plane.  To estimate
the feasibility of observing the invisible supersymmetric
$W\ra\chi\chi$ decays in $e^+e^- \ra W^+W^-$ production processes, we
consider, as an illustrative example, events collected at the LEP 172
GeV run.  Assuming 7\% branching ratio for the supersymmetric $W$
decay modes, one expects both $W$ bosons to decay to standard model
particles in 86.5\% of the cases, both $W$ bosons decaying to
charginos and neutralinos in 0.5\% of the cases, and finally the
signal events defined as one $W$ boson decaying to standard particles
and the other to chargino and neutralino to occur in 13\% of the
cases. The total $WW$ cross section at this energy is $\sim 13$ pb,
which then leads to the signal cross section of the order 1.7 pb.
With the combined integrated luminosity ${\cal L} \sim 4\times 11$
pb$^{-1} =44$ pb$^{-1}$ of the four LEP experiments at $\sqrt{s}=172$
GeV, a total of about 570 $WW$ events have been produced, $i.e.$ 1140
$W$ bosons out of which 80 $W$ bosons are potential candidates for
chargino/neutralino decays if $\mbox{BR}(W\ra\chi\chi)=7$\%.
Therefore 74 signal events with mixed standard and supersymmetric $W$
decays can be expected.

According to the underlying assumptions, the signature of these events
would be a single $W$ boson, $e^+e^-\ra W+$ (no other visible
particle), with the isolated $W$ boson carrying the beam energy.  The
$W$ boson may be tagged in the 2-jet decay mode or, with reduced
branching ratios, in the leptonic $e\nu_e$ and $\mu\nu_{\mu}$ decay
modes. Using selection cuts for single $W$ production consistent with
on-shell $WW$ kinematics, we estimate that in the leptonic tagging
mode ($W\ra e\nu /\mu\nu$) an efficiency at least as large as in the
search for acoplanar lepton pairs, $i.e.$ better than 70\%, and in
the 2-jet tagging mode ($W\ra q\bar{q'}$) an efficiency comparable to
that of the search for $WW\ra \tau\nu q\bar{q'}$, $i.e.$ better than
30\% can be achieved. Combining the cross sections with the
experimental efficiencies etc., it would give rise to $\sim 10$ signal
events in the leptonic, and $\sim 15$ signal events in the hadronic
tagging mode for the LEP172 run.  If the $\mbox{BR}(W\ra\chi\chi)$ is
smaller than 7\%, the expected number of events is reduced
accordingly.

The main backgrounds for the leptonic and 2-jet tagging modes of the
supersymmetric invisible $W$ decays are $WW$ events where one boson
decays leptonically, single $W$ final states $We\nu_e$, and
$q\bar{q}\gamma$ events. In these cases either the lepton or the
photon may escape undetected along the beam pipe giving rise to a fake
"single $W$" signal event.  The cross sections for these background
processes have been obtained with the CompHEP program \cite{boos}
without taking into account hadronization of quarks and detector
effects.  Of course, hadronization of quark jets and smearing due to
the experimental resolution have to be taken into account when an
exact analysis of the signal and background is performed. However,
this needs be done in the experimental analysis, which is beyond the
scope of this paper.

The background from the $WW$ events is small since only in a small
fraction of the $WW\ra Wl\nu_l$ events the lepton is emitted at a
small angle with the beam pipe. The cross section of 0.03 pb is
expected for events with the lepton in a cone of an half-opening angle
$5^o$ around the beam pipe [and 0.02 pb for $WW\ra Wq\bar{q}'$ events
due to the "invisible" SM hadronic decay modes].  The single $W$-boson
production is more difficult to suppress.  An important subprocess in
this channel is the photoproduction process $\gamma e\ra W\nu_e$ with
the Weizs\"acker-Williams photon radiated off the second lepton in the
$e^+e^-$ initial state. This leads to a background cross section of
0.11 pb and 0.32 pb in the leptonic and 2-jet tagging modes,
respectively. Exploiting the special kinematics of the on-shell $WW$
signal process [$i.e.$ the energy $E_i$ of the $W$ decay products is
restricted to the range $ 26\mbox{ GeV} \le E_i \le 62 \mbox{ GeV}$ at
$\sqrt{s}=172$ GeV], the above cross sections can be reduced by 20\%.
The $q\bar{q}\gamma$ final states, with the photon escaping along the
beam pipe, are primarily induced by the radiative return to the $Z\ra
q\bar{q}'$, for which a cross section of 120~pb is predicted
\cite{jadach}.  Even though the cross section is large, it can be
suppressed very efficiently by requiring a cut on the invariant mass
of the two jets, 70~GeV~$\le M_{q\bar{q}'}\le$~90~GeV and the above
cut on jet energies, reducing the value down to 5 pb.  A further cut
on the vector sum of the jet momenta with the beam pipe will reduce
this background to a sufficiently low level.

4. In summary: the special kinematics of on-shell $WW$ production with
2-body $W$ decay, $i.e.$ the invariant mass and the energy conditions,
provide powerful constraints. They can be used to select efficiently
the signal events and to suppress the background processes.
{}From the above estimates of the signal and background,  
both the leptonic and the 2-jet tagging modes seem to be promising
channels for the search for supersymmetric $W$ boson decays at the
level of a few percent. Therefore the analysis of $W$ production in
$e^+e^-$ collisions can be used to exclude part of the area in the
supersymmetry parameter space in which chargino and sneutrino masses
are nearly degenerate -- or to realize this exceptional case
experimentally. The blind spot left in the analysis of chargino pair
production in $e^+e^-$ annihilation can thus partly be closed by
exploiting $WW$ production data.

\noindent
{\bf Acknowledgments :}
We would like to thank W. de Boer for useful comments. Special thanks go 
to D. Zerwas for numerous discussions on the material presented in 
this note. The work of JK is partially supported by the KBN 
grant  2~P03B~180~09.

\vspace{2cm}
\noindent 
{\Large {\bf Figure Captions}}
\begin{enumerate}
\item Contour lines for $\tan\beta=1.5$ in the $[\mu,M_2]$ plane 
along which the branching ratios BR$(W\ra \chi\chi)$ of $W$ decays 
to charginos and neutralinos are 7,5,3 and 1\% (full curves). 
Also shown are the contour lines for the mass bounds 
$m(\chi^0_1)=12$ GeV, $m(\chi^0_2)=45$ GeV and $m(\chi^+_1)=45$ GeV 
(dashed curves).  
\item Blow-up of the positive $\mu$ area in the previous figure 
for the sake of clarity.
\item Contour lines for $\tan\beta=1.5$, $\mu>0$ [upper plot] 
and $\mu<0$ [lower plot] in the 
$[m_{\chi^+_1},m_{\chi^0_1}]$ plane 
along which the branching ratios BR$(W\ra \chi\chi)$ of $W$ decays 
to charginos and neutralinos are 7,5,3 and 1\%. 
\end{enumerate}

\newpage
\begin{figure}[htb]
\centerline{\psfig{figure=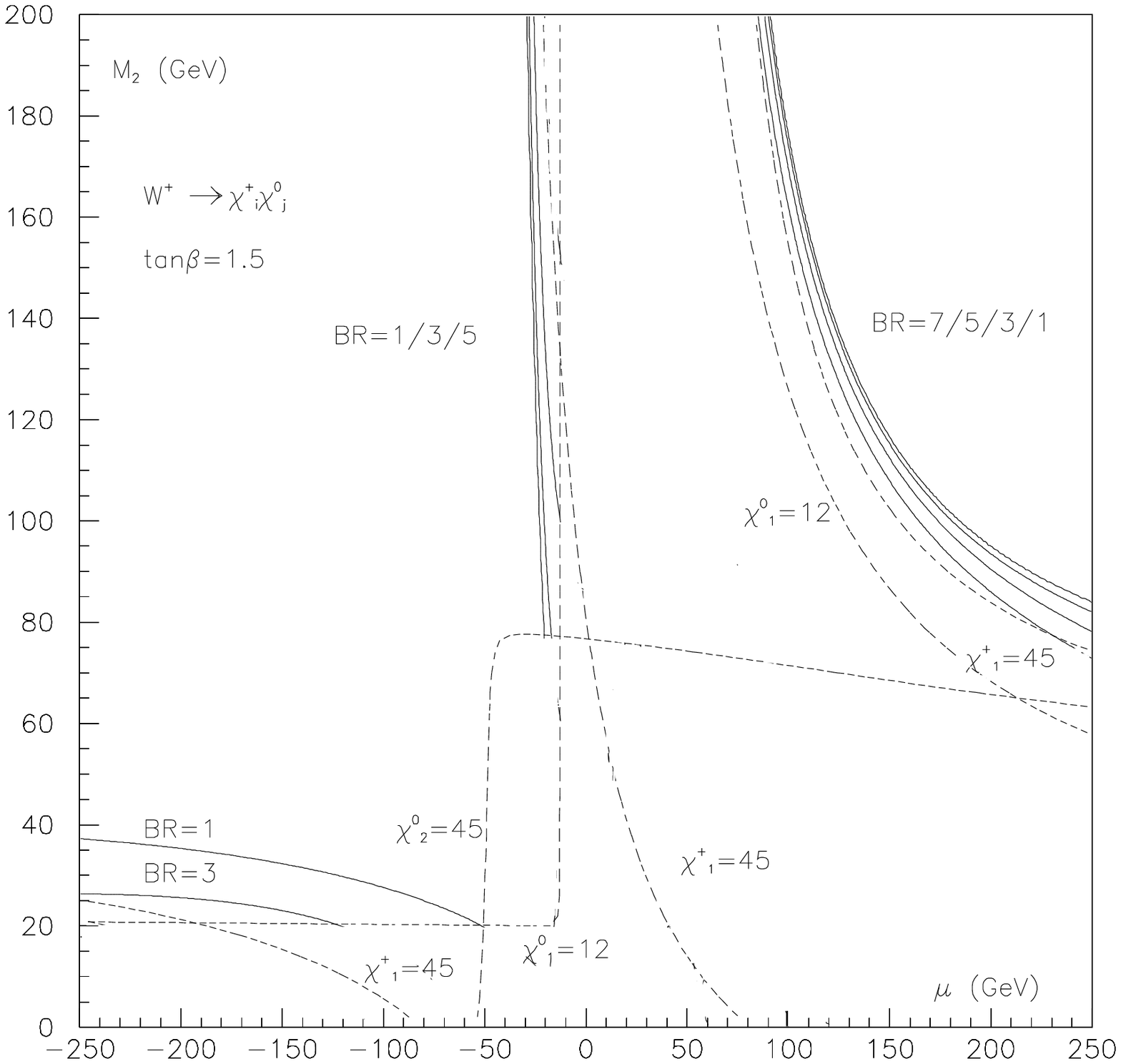,height=17cm,width=17cm}}
\caption{}
\end{figure}
\newpage
\begin{figure}[htb]
\centerline{\psfig{figure=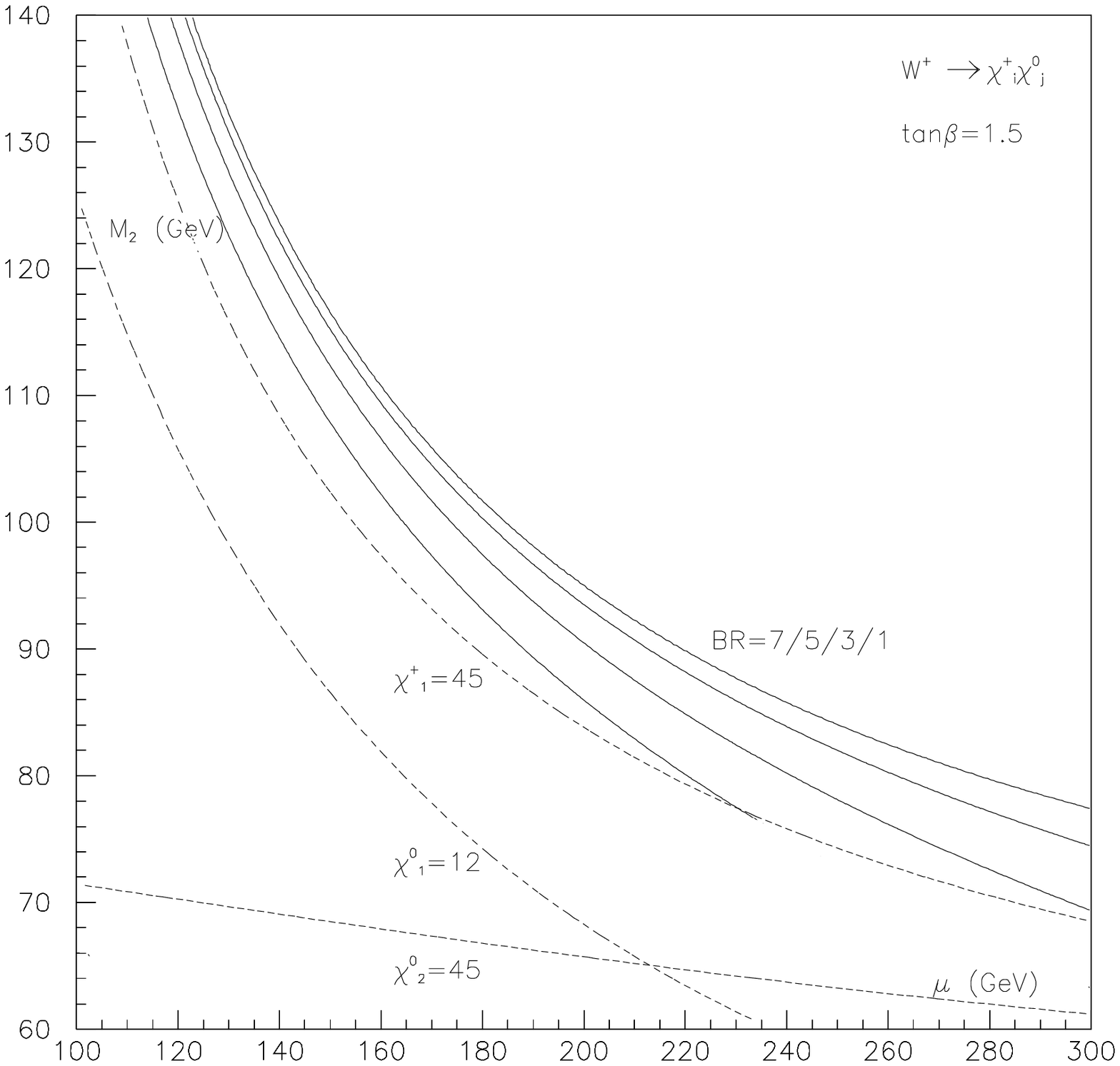,height=17cm,width=17cm}}
\caption{}
\end{figure}
\newpage
\begin{figure}[htb]
\centerline{\hspace{-2cm}\psfig{figure=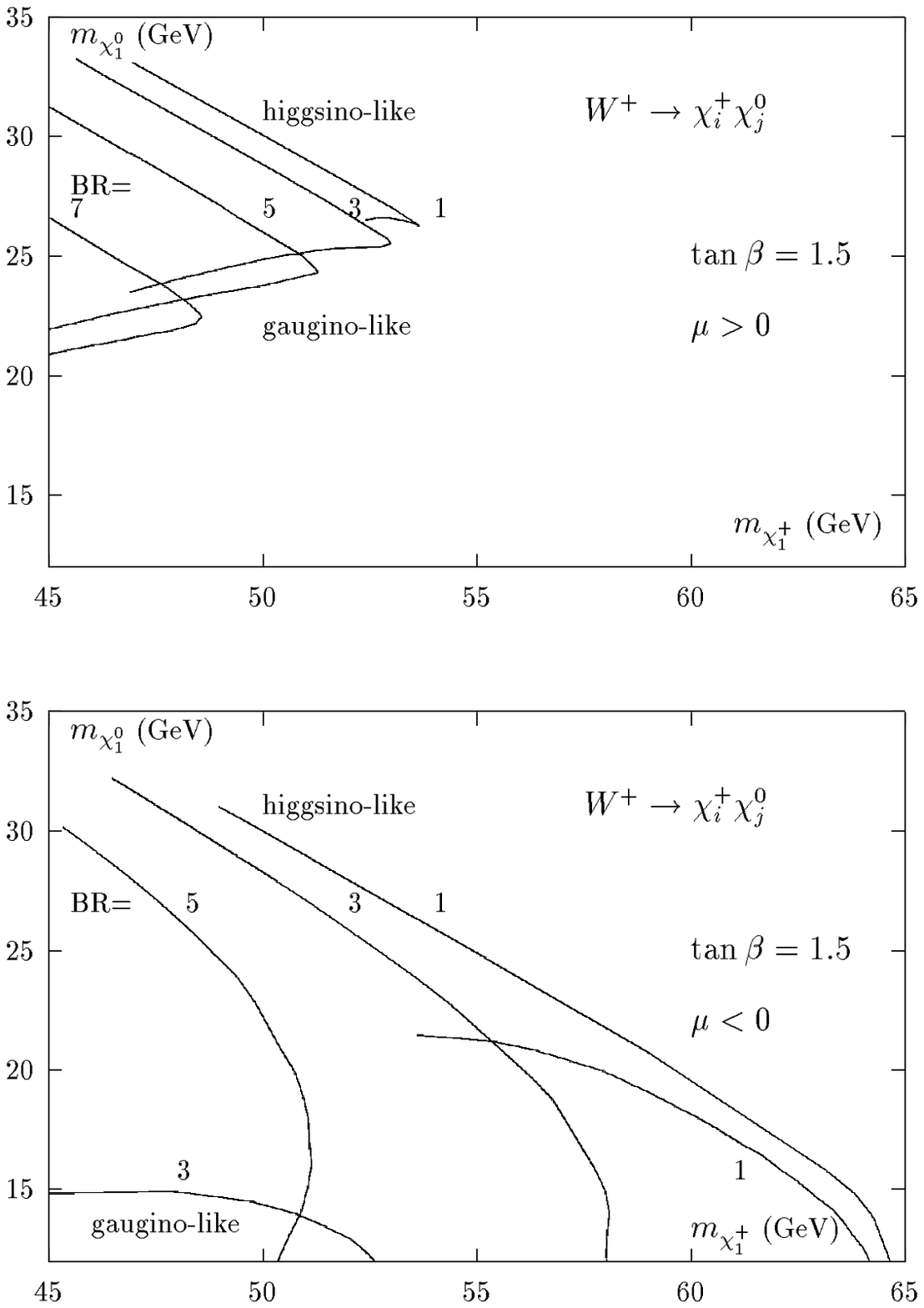,height=22cm,width=17cm}}
\vspace{-5cm}
\caption{}
\end{figure}
\end{document}